# Detection of opposite polarities in a sunspot light bridge: evidence of low-altitude magnetic reconnection

Lokesh Bharti[1*], Thomas Rimmele[2], Rajmal Jain[3], S. N. A. Jaaffrey[1] and R. N. Smartt[2]

1. Department of Physics, University College of Science,
Mohanlal Sukhadia University,
Udaipur, 313001, India
(E-mail-lokesh_bharti@yahoo.co.in)
2. U. S. National Solar Observatory, Sunspot, Sacramento Peak,
New Mexico, 88349, USA
3. Department of Space (Govt. of India), Physical Research Laboratory,
Navrangpura, Ahmedabad, 380 009, India

**ABSTRACT**

A multi-wavelength photometric analysis was performed in order to study the sub-structure of a sunspot light bridge in the photosphere and the chromosphere. Active region NOAA 8350 was observed on 1998 October 8. The data consist of a 100 min time series of 2D spectral scans of the lines FeI 5576 Å, Hα 6563 Å, FeI 6302.5 Å, and continuum images at 5571 Å. We recorded line-of-sight magnetograms in 6302.5 Å. The observations were taken at the Dunn Solar Telescope at U. S. National Solar Observatory, Sacramento Peak. We find evidence for plasma ejection from a light bridge followed by Ellerman bombs. Magnetograms of the same region reveal opposite polarity in light-bridge with respect to the umbra. These facts support the notion that low altitude magnetic reconnection can result in the magnetic cancellation as observed in the photosphere.

Subject headings: Sun: sunspots, Sun: magnetic fields, Sun: photosphere, Sun: chromosphere

## 1. INTRODUCTION

Large sunspots often have a complex structure and vary on spatial and temporal scales. During the lifetime of a sunspot its umbra may be crossed by one or more narrow, bright bands known as "light bridges". One of the early signature of imminent sunspot breakup is the formation of a light bridge: an intrusion, or lane of bright material that cuts across the umbra of a sunspot ( Bray & Loughhead 1964). Light Bridges "show a great diversity in shape, size and brightness" (Bray & Loughhead 1964, p. 89). According to Vázquez (1973), the formation of so-called "photospheric light bridge" in sunspot or pores is a result of the decay of the spot and restoration of the granular surface as precursors of the imminent decay of active region. The presence of umbral bright features (umbral dots and light-bridges) can be explained theoretically either in terms of cluster model, where the umbra is formed by tight bundle of isolated thin flux tubes, separated by field-free columns of hot plasma (Parker 1979 a, b; Choudhuri 1986), or in terms of magneto convection taking place in a coherent but homogeneous large flux tube (Knobloch & Wiess 1984, Weiss et al. 1990, 1996, Blanchflower et al. 1998).

The internal structure of light bridges can be granular or filamentary. According to their width, light bridges are classified by Sobotka *et al*. 1993, 1994. A strong light bridge, which separates large umbral fragments, is further distinguished according to the fine structure observed within it: penumbral (strong) light bridges exhibit filamentary fine structure and separate umbral fragments of opposite polarity, while photospheric (strong) light bridges show fine structure similar to photospheric granulation (Sobotka et al. 1994 and references therein, Rimmele, 1997). Muller (1973, 79), studied strong penumbral light bridges and concluded that the morphological and physical properties of filamentary structures in light bridges are similar those in the penumbra. The faint light bridge most likely consists of a chain of umbral dots which gives the appearance of a faint narrow lane within the umbra (Muller 1979). Observations from Swedish 1-m Solar Telescope (Scharmer et al. 2003a, SST) reveal new structure of light bridges. Berger & Berdyugina (2003) reported that dark central lanes are common feature of strong light bridges, running along the length of bridge. Light bridges are segmented with bright grains along length and separated by narrow dark lane oriented perpendicular to the length of the bridge. These bright grains show unidirectional flow along light bridge length. Lites et al. (2004) found that segmented structure of light bridges running through sunspots and pores are raised above the dark umbral background and some light bridges that extend into penumbra do not have distinct bright grains or dark lanes and they resemble more the elongated bright filamentary structure seen in the penumbra.

Rüedi et al. (1995) reported a downflow on the order of 1.5 km/s in a photospheric light bridge while Beckers & Schröter (1969) find blueshift. Le ka (1997) studied an ensemble of 11 light bridges and found that light bridges randomly show either redshifts or blueshifts. From time series of 2D scan of Fe I 5576 Å line Schleicher et al. (2003) reported slight blue shift, interrupted by short events of strong upflow or downflow within light bridge. A positive correlation between the brightness and up flow velocities was reported by Rimmele (1997), who showed that velocity cells in the light bridge have a longer lifetime (20-30 minutes compared to 7 minutes for granulation). These bright grains having tendency to reappear in the same location. Rimmele interpreted his observations as evidence for the magneto convective origin of photospheric light bridges. However, the character of the convection differs significantly from the normal granular convection observed in the quiet photosphere (Sobotka et al. 1993, 94, Rimmele 1997). Rimmele (1997) did not observed the narrow, dark lanes containing strong downflows in granular convection in the light bridge. On basis of similarities in the bisector asymmetry of the Fe I 5434 Å line observed in a light bridge and in the quiet photosphere Sobotka et al. (1994) suggest a convective origin of the light bridge. Light bridges have been found to have the same magnetic polarity as that of the sunspot umbra, while exhibiting a systematic reduction of the magnetic field strength compared with the surrounding umbra. In addition the field in the light bridge is more inclined with respect to the vertical direction (Beckers & Schröter(1969), Abdusamatov (1970), Kneer (1973), Wiehr & Degenhardt (1993), Rüedi et al. (1995) and Leka (1997)).

Observations of light-bridges seen in the chromosphere and corona by Roy (1973) and Asai, Ishii & Kurokawa (2001) respectively show the remarkable phenomena of "plasma ejection" or "surge" activity. However, the origin of such ejection in the context of magnetic configuration is still not clear. Observations reported so far are from within regions of equal polarity with no reported observations of intrusions of opposite polarity fields accompanied by chromospheric observations of light bridges. We report for the first time in this letter the observations of an unusual case where opposite polarity from the surrounding umbra is seen in the light-bridge and associated with observed plasma ejection.

## 2. OBSERVATIONS AND DATA REDUCTION

The observations for this investigation were made at the U. S. National Solar Observatory, Sacramento Peak on 1998 October 08 using the 76-cm Dunn Solar Telescope from 14:05 to 15:45 UT. The main instrument used for these observations was the Universal Birefringent Filter (UBF). The UBF is a tunable Lyot filter with a passband that varies between 180 and 250 mÅ as a function of wavelength. We used 250 mÅ filter width for our observations. The surge activities along the light-bridge in NOAA AR 8350 (N19$^o$, W10$^o$) were observed with the tuning at 5571.9 Å continuum, then at the blue wing of the Fe I 6302.5 Å with an offset of 80 mÅ from line centre to obtain line-of-sight (LOS) magnetograms. A quarter-wave plate mounted in front of the UBF enabled us to record left and right hand circular polarization (LCP and RCP) filtergrams, respectively. The UBF was then tuned to line center and into the red and blue wings of Fe I 5576 Å, Fe I 6302 Å and Hα 6563 Å at different line positions. According to Altrock et al. (1975), the core of the Fe I 5576 Å line, which forms at an altitude of about 320 km, was used to obtain Dopplergrams at different heights in the solar atmosphere. However the height of formation of this line changes slightly in active regions (Bruls, Lites, & Murphy 1991). The Fe I 5576 Å line has the advantage of being a "nonmagnetic" line; i.e., the effective Landé factor is $g$=0, providing a clean Doppler signal without any cross talk from the magnetic field. We calibrate the magnetogram and dopplergram described by Berger & Title (2001) and Rimmele (2004). The noise floor in the magnetograms was determined from the FWHM of a Gaussian fit to the histogram of the calibrated magnetogram (see Fig 5 of Berger & Title 2001). The noise level typically ranges from about 115 to 180 gauss.

In addition to the UBF data, we recorded G-band images simultanously, using a 1nm wide passband interference filter centered at the CH-band head at 4305 Å. The UBF filtergram sequences were recorded with a Xedar 1k x 1k CCD camera that enabled a pixel resolution of 0´´.117/pixel. The G-band observations were made with a Thomson 1k x 1k CCD camera achieving almost the same pixel resolution of about 0´´.119/pixel. The exposure time for the G-band camera was 12 ms and for UBF camera varied from 100 - 300 ms depending upon the wavelength of the observation. The UBF sequence cycle contains 28 filtergrams at a cadence of 5 s simultaneously with G-band images. Hence the time to complete one observing cycle was 2.5 min. We collected data over a period of about 100 minutes from which the best images were selected for further processing and analysis.

After the standard flat and dark current correction, we applied destretch algorithm to correct residual differential motions visible across the extended FOV, caused by seeing effects to the G-band images and UBF filtergrams sequences. To align G-band images and UBF filtergrams, we scaled down UBF filtergrams image scale to that of G-band images using a cubic spline interpolation.

## 3. RESULTS

### 3.1 Continnum 5571 Å observations

As shown in Figure 1(left), the observed light bridge can be classified as a strong penumbral light bridge, which divides the umbra into two parts. A well developed penumbra on the two opposite sides is clearly visible. Sunspot fine structures such as umbral dots can be seen in both umbral fragments and a faint light bridge is visible in the upper umbra. The light bridge bend in the middle. A dark central lane is also visible in one half of it as reported by Berger & Berdyugina (2003) and Lites et al. (2004). Fine structures can be seen in the light bridge. Some of them have a penumbral structure: they show elongated bright grains lined up into filaments as in the penumbra and their brightness is close to that of penumbral filaments. This light bridge shows such a 'penumbral light bridge', which crosses the main umbra of a sunspot. Other structures appear to consist of bright grains similar to abnormal granules found around sunspot (Sobotka *et al*., 1994). These are seen in higher numbers in the left region of the light bridge. Adjacent to the right region a well developed penumbra is visible while adjacent to the left region there is no penumbra. Instead abnormal granulation is visible there.

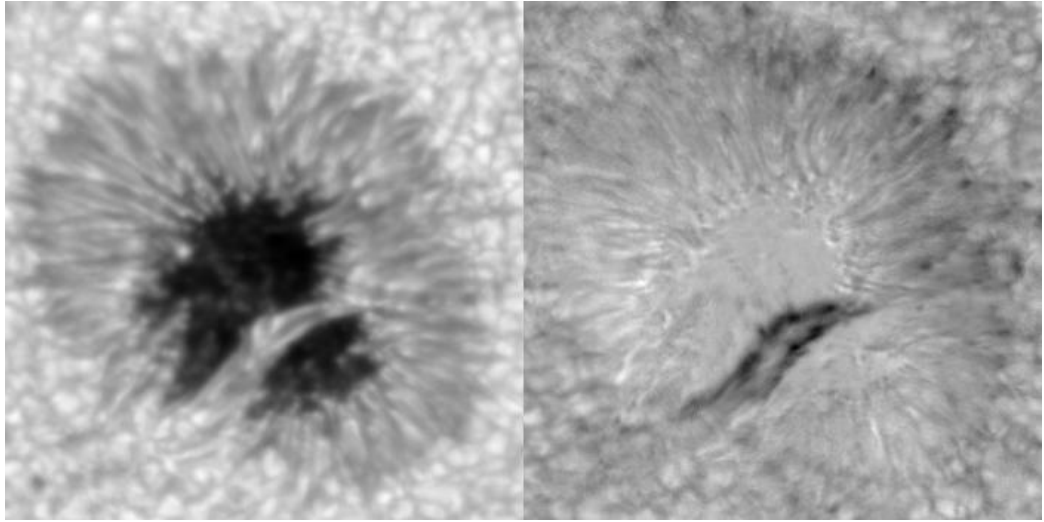

Figure 1. 5571 Å image of sunspot (left) and 5576 Å dopplergram (right): fine structures of sunspot and light bridge are clearly visible within the line of sight velocity.

### 3.2 Dopplergram

Figure 1(right) shows one of the best Dopplergram obtained. Dark indicates downflow and bright indicates upflow. We chose the darkest part of umbra as a zero point reference for the LOS velocity. At the edge of light bridge a strong down flow up to 0.5-1.4 km/s is observed while in the middle part of the light bridge an upflow around 0.1-0.2 km/s is measured. The upflow observed in quiet sun granulation in range 0.2- 0.8 km/s while downflow in range 0.2- 0.5 km/s. Hence upflow in light bridge is significantly lower then quiet sun granulation while the strong downflow observed at the edges of light bridge arms is 2-3 times larger in magnitude than that of downflowing cool plasma in intergranular lanes.

### 3.3 Hα observations

Movies of Hα line centre and Hα-700 mÅ filtergrams reveal the presence of Ellerman bombs (Ellerman 1917) at different locations of the light bridge and adjacent to left part. The mean relative peak intensity of these bright patches with respect to the quiet chromosephere is 1.67. These Ellerman bombs are accompanied by dark surges. These conspicuous activities can be seen more clearly in Hα-700 mÅ filtergrams (Figure2). Figure 2(a) shows a dark surge (marked by the arrow) that ejected from the upper right area of the light bridge and projected over the adjacent penumbra. We observed repeated ejections from the same location as shown in Figures 2 (a-h). Figures 2 (f-h) show another surge ejected from the lower left region of the LB that was accompanied by Ellerman bombs. The measured projected velocity of the ejection is 4.5 km/s. At 15:14:20 UT in Figure 2 (e) Ellerman bombs are clearly seen at lower left region before this surge. During our observing time we also observed surges followed by Ellerman bombs at the edge of lower left region of light bridge as shown in Figure 2 (a-i). In Figure 3 we marked location of this ejection site by a arrow to show system of two opposite polarities.

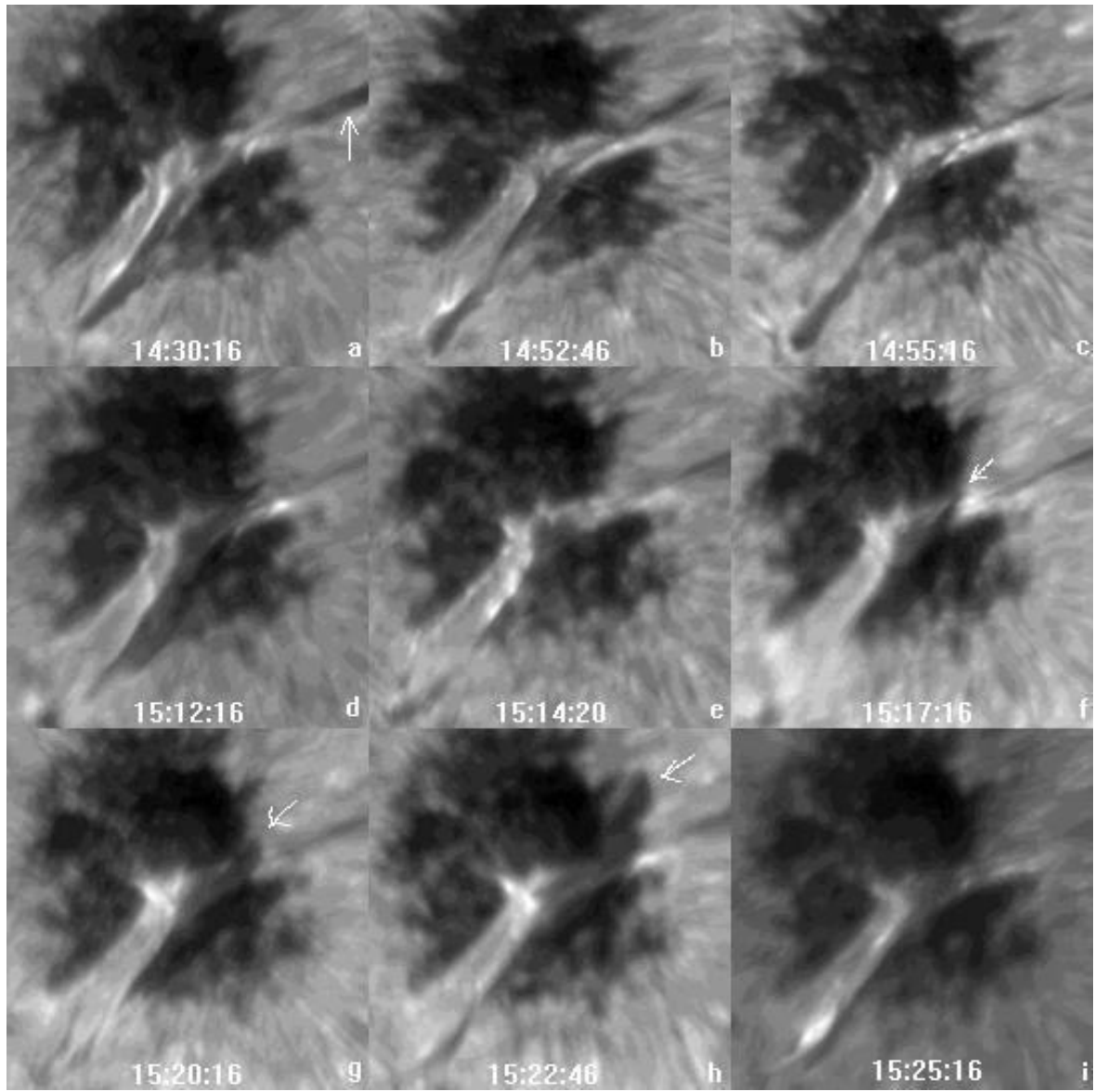

Figure2. Ellerman bombs and surges over the light bridge and near to lower left arm edge in Hα-700 mÅ.

## 3.4 Magnetic configuration:

The continuous mass ejection from light bridge was considered to be an evidence of emerging flux (e.g. Kurokawa & Kawai 1993). Our high resolution magnetograms (Figure3) showed, for the first time, a remarkable feature of opposite polarity in the lightbridge with respect to the umbra. Evidence of magnetic flux cancellation can be seen at the edge of lower left region. This ejection site shown by the arrow in Figure 3 where opposite polarity flux of light bridge cancelled with positive flux and accompanied by Ellerman bombs and surges is observed in the Hα line. Magnetic structure at this location change significantly between Figure 3(a) and 3(b). Ellerman bombs and surges along length of the light bridge may be due to flux cancellation between opposite polarity light bridge and positive polarity flux of umbra. However we could not make quantitative measurement of flux cancellation since our magnetogram suffers from saturation at higher magnetic field strength.

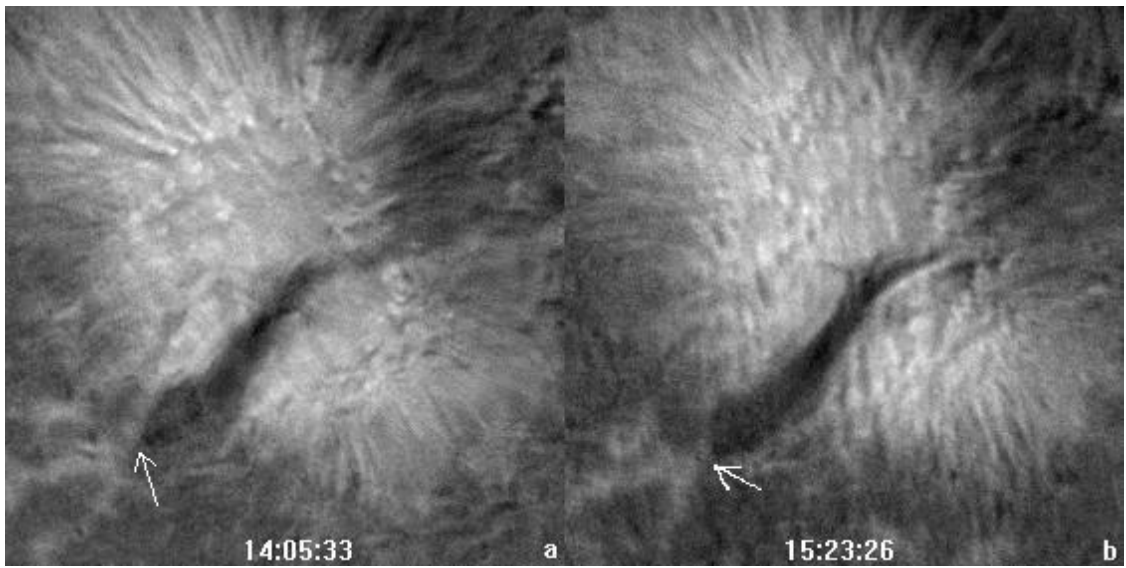

Figure 3. Snap shots of sunspot magnetogram. The opposite polarity in light bridge is clearly visible. The gray scale ranges from – 350 to 780 gauss.

Since the sunspot is near disk centre projection effects are minimized. We compared our magnetograms with the same day Kitt-Peak magnetogram of the same sunspot and found that Kitt-Peak magnetogram do not show opposite polarity in the light bridge, however our magnetogram does. Our magnetogram has 9.6 times the spatial resolution of Kitt Peak and therefore it is not surprising that this feature is not seen in the Kitt-Peak magnetogram. To verify that the opposite polarity signal is real, we examine the strength of the signal relative to the noise and find that it is 244% above the noise level. In addition, the signal is seen in successive magnetogram images and it evolves continuously, as would be expected of a real polarity structure. Therefore we believe that the opposite polarity feature is real and is not an artifact. Kitt-Peak magnetogram does not see the feature because it is smeared out by the low spatial resolution of their instrument.

# 4. DISCUSSION AND CONCLUSIONS

It is widely believed that the bright knots and the plasma ejection observed in the chromosphere are driven by emerging flux that reconnects successively with the ambient fields. Wang & Shi (1993) and Yoshimura *et al.* (2003) argued that magnetic reconnection may lead to flux cancellation because the cancellation always takes place at the interface region between different magnetic field systems. The frequent mass ejections observed in the chromosphere above the observed lightbridge could be interpreted as the signature of such reconnection events occurring in the upper photosphere and lower chromosphere, triggered by magnetic flux cancellation that occur between the light-bridge and umbra. However it is not straightforward to quantify formation height in the chromosphere from H$\alpha$ observations. According to Socas-Navarro & Uitenbroek (2004) this is due to the very core of the line is optically thick in the chromosphere and main contribution to wings, from most of the upper photosphere and lower chromosphere. Height of formation calculations are typically based on quiet Sun plan-parallel model atmospheres which are not applicable in a sunspot light bridge. However we believe that they help to support our hypothesis that magnetic reconnection is taking place in the photosphere leading to an upward propagation heat source and mass flow in the light bridge atmosphere. Recent studies have shown that the low altitude reconnections give evidence to be the direct cause of the ubiquitous micro flares or Ellernman bombs observed in emerging flux regions (Chae et al. 1999, Mandrini et al. 2002; Schmieder et al. 2002, Liu & Kurokawa 2004). The low altitude reconnection process effectively helps to inject dense plasma into the upper atmosphere. It is evident from our magnetogram that the polarity of light bridge and umbra are opposite and therefore continuous magnetic field cancellation may be a dominant process to trigger the mass ejection seen in our chromospheric filtergrams.

In order to understand plasma ejection from light bridges and other magnetic fields, of complex topology of magnetic field, high resolution multi wavelength and spectropolametric observation are needed. This could be possible from Solar-B space mission and from Diffraction Limited Spectropolarimeter (DLSP) (Sankarasubramanian *et al.*, 2004) at DST.

# ACKNOWLEDGEMENT

Thanks are due to an anonymous referee for valuable suggestions and critical comments leading to improvement of the manuscript. The authors are thankful to the NSO observing staff for their help in our two week observing program. We are grateful to Dr. Han Uitenbroek (NSO) for much valuable discussions on formation heights of spectral lines used for this analysis and Prof. P. Vanketkrishnan (USO) and Dr. Sankarsubranian (NSO) for valuable discussions and suggestions. We are also thankful to Dr. K.B. Joshi and Chandan Joshi for their help in preparing the manuscript. Lokesh Bharti is thankful to the 22$^{nd}$ NSO workshop organizers for providing travel support and local hospitality during his stay at Sac Peak where he did substantial work for this investigation. This research is supported by Bal Shiksha Sadan Samiti, Udaipur.